\documentclass{article}
\usepackage{tikz, subcaption, graphicx, braket, amsmath, cite, amssymb, color}
\usepackage[colorlinks = true]{hyperref}
\usepackage[margin = 2cm]{geometry}

\DeclareMathOperator{\bin}{bin}

\title{Qudit States in Noisy Quantum Channels}
\author{Supriyo Dutta\thanks{Email: \texttt{dosupriyo@gmail.com}} \\ Department of Mathematics, National Institute of Technology Agartala,\\ Jirania , West Tripura, India -  799046 \vspace{.25 cm} \\ Subhashish Banerjee\thanks{Email: \texttt{subhashish@iitj.ac.in}} and Monika Rani\thanks{Email: \texttt{rani.2@iitj.ac.in}} \\ Department of Physics, Indian Institute of Technology Jodhpur \\ Jodhpur, Rajasthan, India - 342037.} 
\date{\today}

\begin{document}

\maketitle

\begin{abstract}
In this work, we analyze a number of noisy quantum channels on a family of qudit states. The channels studied are the dit-flip noise, phase flip noise, dit-phase flip noise, depolarizing noise, non-Markovian Amplitude Damping Channel (ADC), dephasing noise, and depolarization noise. To gauge the effect of noise, the fidelity between the original and the final states is studied. The change of coherence under the action of noisy channels is also studied, both analytically and numerically. Our approach is advantageous as it has an explicit relation to the original approach to the multi-qubit hypergraph states.
\end{abstract}

\section{Introduction}
	
	Technological and theoretical advances are making qudit states indispensable in quantum information and computation. Quantum algorithms represent a prominent application within the field of modern quantum information theory, offering the potential for computational acceleration that classical systems are unlikely to ever achieve. A renowned approach for implementing quantum algorithms involves the creation of exceptionally entangled graph states of a particular type. Hypergraph states, also known as multipartite entangled states or high-order entangled states, are quantum states that extend the concept of entanglement beyond pairwise correlations typically found in Bell states or graph states. They offer a platform to generalize the ideas that were initially developed for qubit states. Thus, for example, qudit states have found application in quantum teleportation \cite{zhang2019quantum, dong2008controlled, fonseca2019high},  quantum computing \cite{wang2020qudits, kiktenko2020scalable, luo2014universal}, quantum walk \cite{giordani2019experimental, li2019implementation, giordani2021entanglement}, and quantum state transfer \cite{paz2009perfect, li2021discrete, jafarizadeh2009perfect}. Quantum systems are invariably affected by noise due to interaction with the ambient environment \cite{banerjee2018open}. Therefore, a study of the dynamics of qudit states evolving under noisy conditions is a relevant issue, which we take up here.

Qudits are higher dimensional generalizations of qubits that are increasingly gaining importance in several areas of quantum science and technology \cite{adepoju2017joint, falaye2016jrsp}. Noise is an always unavoidable phenomena in any physical system. Particularly quantum noises have very special features with their effects being characterized through non-reversible operators. In this paper, we focus on studying how noise affects quantum states.
In order to investigate the impact of noise on a state, the characteristics of the corresponding noisy quantum channel should be understood. The quantum channels are represented by appropriate Kraus operators. Fidelity is a useful diagnostic for this. The quantum channels which we study are dit-flip noise, phase flip noise, dit-phase flip noise, depolarizing noise, ADC (non-Markovian noise), non-Markovian dephasing noise, and non-Markovian depolarization noise \cite{utagi2020temporal,shrikant2018non}. These channels were originally defined to apply to qubits. The dit-flip noise, phase flip noise, dit-phase flip noise, and depolarizing noise were generalized to apply on qudit states 
in \cite{fonseca2019high}. Following this direction, we generalize ADC (non-Markovian noise), non-Markovian dephasing, and non-Markovian depolarization noise on 
qudits. The analytical expression of fidelity, between the original and the final states, is calculated for each of these channels. This serves 
to gauge the impact of the noise, in consideration, on the quantum state. Coherence is the root cause of most of the intriguing features of quantum mechanics.
We note that the coherence of these states is consistently decreasing after the application of these channels. 

This article is distributed as follows. Preliminary concepts are laid down in Section 2. 
We begin Section 3 with the Weyl operator formalism that is essential to design qudit channels. We dedicate different subsections to different channels. We discuss unital as well as non-unital channels, both Markovian and non-Markovian. In every subsection, we discuss the fidelity between the original and the final state as well as the change in coherence. We generalize these quantum channel at the end of this section. Then we make our conclusion. In an appendix, we draw an analogy of the family of qudit states discussed with quantum hypergraph states.

	\section{Preliminaries}

		We consider number states $\ket{0}, \ket{1}, \dots \ket{N-1} \in \mathbb{C}^N$ to define a family of quantum states 
		\begin{equation}\label{qudit_hypergraph_state}
			\ket{G} = \frac{1}{\sqrt{N}} \sum_{i = 0}^{N - 1} (-1)^{g(i)}\ket{i},
		\end{equation}
		where, $g: \{0, 1, 2, (N - 1)\} \rightarrow \{0, 1\}$ is a Boolean function. The qudit states $\ket{G}$ exhibit an analogy with the well-known quantum hypergraph states, which is brought out in the Appendix. The density matrix of the state can be expressed as
\begin{equation}\label{qudit_density_in_full_form}
\rho = \ket{G}\bra{G} = \sum_{i = 0}^{N  -1} \sum_{j = 0}^{N - 1} \frac{(-1)^{g(i) + g(j)}}{N} \ket{i}\bra{j}.
\end{equation}

To date, there are a number of techniques for investigating the evolution of a quantum state in the context of open quantum systems \cite{banerjee2018open}. The Kraus operator formalism \cite{choi1975completely, kraus1974operations, kraus1983states} finds a prominent place in this context. In this method, the evolution of the quantum state $\rho$ is modeled by a set of trace-preserving maps $\{E_k: k = 1, 2, \dots\}$. The final state is represented by
\begin{equation}\label{Kraus_operator}
\Lambda(\rho) = \sum_k E_k \rho E_k^\dagger, ~\text{where}~ \sum_k E_k^\dagger E_k = I
\end{equation}
is the identity operator. 

One of the central issues in the investigations of open quantum systems is the dynamics of decoherence. It is concerned with the evolution of quantum coherence, which is 
particularly important for quantum information and computation to carry out tasks otherwise impossible within the realm of classical physics. The $l_1$ norm of coherence is an important measure of quantum coherence in a state, which we can 
analytically calculate \cite{baumgratz2014quantifying}. Given a density matrix $\rho = (\rho_{i,j})$, the $l_1$ norm of coherence is defined by
\begin{equation}
C_{l_1}(\rho) = \sum_{i \neq j} |\rho_{i,j}| = \sum_{i = 0}^{N - 1} \sum_{\substack{j = 0 \\ j \neq i}}^{N - 1} |\rho_{i,j}|.
\end{equation}
From the equation (\ref{qudit_density_in_full_form}) we can see that the $(i,j)$-th entry of the density matrix $\rho$ is $\rho_{i,j} = \frac{(-1)^{g(i) + g(j)}}{N}$. Therefore $|\rho_{i,j}| = \frac{1}{N}$. Hence, the $l_1$ norm of coherence of any member of these states is
\begin{equation}\label{coherence_of_hypergraph_state}
\begin{split}
C_{l_1}(\rho) & = \sum_{i \neq j} |\rho_{i,j}| = \sum_{i = 0}^{N - 1} \sum_{\substack{j = 0 \\ j \neq i}}^{N - 1} |\rho_{i,j}| = \frac{N^2 - N}{N} = N - 1.
\end{split}
\end{equation}
Another important tool to gauge the impact of open system effects, on the quantum states is to study the evolution of fidelity, quantifying 
the closeness of two states, under the noise in consideration. This will be taken up here for the cases studied.

\section{Noisy quantum channels in higher dimension}

In this section, we generalize a number of quantum channels for qudit states. The mathematical preliminaries, laid down here will be used subsequently in this paper. The Weyl operators were first introduced in the context of quantum teleportation \cite{bennett1993teleporting}. This is well-studied in the context of quantum computation and information \cite{bertlmann2008bloch, fonseca2019high, narnhofer2006entanglement, baumgartner2006state, baumgartner2007special}. For an $N$ dimensional qudit system there are $N^2$ Weyl operators $\hat{U}_{r, s}$, such that \cite{bertlmann2008bloch}
\begin{equation}\label{Weyl_operator_in_general}
\hat{U}_{r, s} = \sum_{i = 0}^{N - 1} \omega_{N}^{i r} \ket{i}\bra{i \oplus s} ~\text{for}~ 0 \leq r, s, \leq (N - 1),
\end{equation}
where $\omega_{N} = \exp(\frac{2 \pi \iota}{N})$ is the primitive $N$-th root of unity, and ``$\oplus$" denotes addition modulo $N$. Clearly, $\hat{U}_{0, 0} = I_{N}$, the identity matrix of order $N$. Also, $\hat{U}_{r, s}^\dagger \hat{U}_{r, s} = \hat{U}_{r, s} \hat{U}_{r, s}^\dagger = I_N$, that is $\hat{U}_{r, s}$ is a unitary operator for all $r$ and $s$. Applying $\hat{U}_{r, s}$ on state $\ket{G}$, in equation (\ref{qudit_hypergraph_state}) we have
\begin{equation}\label{Weyl_on_G}
\begin{split}
\hat{U}_{r, s} \ket{G} & = \sum_{i = 0}^{N - 1} \omega_{N}^{i r} \ket{i}\bra{i \oplus s } \left[\frac{1}{\sqrt{N}} \sum_{j = 1}^{N - 1} (-1)^{g(j)}\ket{j}\right] = \frac{1}{\sqrt{N}} \sum_{i = 0}^{N - 1} (-1)^{g(i \oplus s)} \omega_{N}^{i r} \ket{i}.
\end{split}
\end{equation}
As $\omega = \exp\left(\frac{2 \pi \iota}{N}\right)$, we have $\overline{\omega^j} = \omega^{-j}$. Using this, the application of $\hat{U}_{r, s}$ on the density matrix $\rho$ in equation (\ref{qudit_density_in_full_form}) is seen to be
\begin{equation}\label{Weyl_on_G_density}
\begin{split}
& U_{r, s} \rho U_{r, s}^\dagger = U_{r, s} \ket{G}\bra{G} U_{r, s}^\dagger = \frac{1}{N} \sum_{i = 0}^{N - 1} \sum_{j = 0}^{N - 1} (-1)^{g(i \oplus s) + g(j \oplus s)} \omega_{N}^{(i - j)r} \ket{i}\bra{j}.
\end{split}
\end{equation}
The following expressions will be useful in the calculations below. From equation (\ref{Weyl_on_G}) we have
\begin{equation}\label{Fidelity_half}
\braket{G| U_{r,s} | G} = \frac{1}{\sqrt{N}} \sum_{i = 0}^{N - 1} (-1)^{g(i \oplus s)} \omega_{N}^{ir} \braket{G | i} = \frac{1}{\sqrt{N}} \sum_{i = 0}^{N - 1} (-1)^{g(i \oplus s)} \omega_{N}^{ir} \frac{(-1)^{g(i)}}{\sqrt{N}}.
\end{equation}
Also,
\begin{equation}\label{Fidelity}
\braket{G| U_{r,s} \rho  U_{r,s}^\dagger | G} = \braket{G| U_{r,s} | G} \braket{G| U_{r,s}^\dagger | G} = \frac{1}{N^2} \left| \sum_{i = 0}^{N - 1} (-1)^{g(i \oplus s) + g(i)} \omega_{N}^{ir} \right|^2.
\end{equation}
We now apply different noisy channels on the state $\rho$ in equation (\ref{qudit_density_in_full_form}).		

\subsection{Dit-flip noise} 

The dit-flip noise is a generalization of bit-flip noise. It flips the state $\ket{i}$ to the state $\ket{i \oplus 1}, \ket{i \oplus 2}, \dots \ket{i \oplus N - 1}$ with probability $p$. The associated Kraus operators are 
\begin{equation}
\hat{E}_{0, s} = \begin{cases} \sqrt{1 - p} I_{N} & ~\text{when}~ r = 0, s = 0; \\ \sqrt{\frac{p}{N - 1}} U_{0, s} & ~\text{when}~ r = 0, 1 \leq s \leq (N - 1). \end{cases}
\end{equation}
The new state after applying the dit-flip operation is
\begin{equation}
\rho(p) = \sum_{s = 0}^{N - 1} E_{0, s} \rho E_{0, s}^\dagger = (1 - p)\rho + \frac{p}{N - 1} \sum_{s = 1}^{N - 1}  U_{0, s} \rho U_{0, s}^\dagger.
\end{equation}			
Applying equations (\ref{qudit_density_in_full_form}) and (\ref{Weyl_on_G_density}), the final state is
\begin{equation}\label{final_state_after_dit_flip}
\begin{split}
\rho(p) & = (1 - p) \rho + \frac{p}{N(N - 1)} \sum_{s = 1}^{N - 1} \sum_{i = 0}^{N - 1} \sum_{j = 0}^{N - 1}  (-1)^{g(i \oplus s) + g(j \oplus s)} \ket{i} \bra{j} \\
& = \sum_{i = 0}^{N - 1} \sum_{j = 0}^{N - 1} \left[(1 - p)\frac{(-1)^{g(i) + g(j)}}{N} + \frac{p}{N(N - 1)} \sum_{s = 1}^{N - 1} (-1)^{g(i \oplus s) + g(j \oplus s)} \right] \ket{i} \bra{j}.
\end{split}
\end{equation}
Clearly, the $(i,j)$-th element of $\rho(p)$ in equation (\ref{final_state_after_dit_flip}) is given by 
\begin{equation}
\rho_{i,j} = (1 - p) \frac{(-1)^{g(i) + g(j)}}{N} + \frac{p}{N(N - 1)} \sum_{s = 1}^{N - 1} (-1)^{g(i \oplus s) + g(j \oplus s)}.
\end{equation} 
Hence, the $C_{l_1}$ norm of coherence is
\begin{equation}
C_{l_1}(\rho(p)) = \sum_{i = 0}^{N - 1} \sum_{\substack{j = 0 \\ j \neq i}}^{N - 1} |\rho_{i,j}| = \sum_{i = 0}^{N - 1} \sum_{\substack{j = 0 \\ j \neq i}}^{N - 1} \left|(1 - p) \frac{(-1)^{g(i) + g(j)}}{N} + \frac{p}{N(N - 1)} \sum_{s = 1}^{N - 1} (-1)^{g(i \oplus s) + g(j \oplus s)} \right|,
\end{equation}
which depends on different choices of the state $\rho$. Clearly,
\begin{equation}
C_{l_1}(\rho(p)) \leq \sum_{i = 0}^{N - 1} \sum_{\substack{j = 0 \\ j \neq i}}^{N - 1} \left[ \frac{(1 - p)}{N} + \frac{p}{N} \right] = \sum_{i = 0}^{N - 1} \sum_{\substack{j = 0 \\ j \neq i}}^{N - 1} \frac{1}{N} = \frac{N(N - 1)}{N} = (N - 1),
\end{equation}
which is the $l_1$ norm of coherence of the initial state, see equation (\ref{coherence_of_hypergraph_state}). Therefore, the coherence decreases on the application of the dit-flip noise on $\rho$.  

Now we calculate the fidelity between the initial state $\rho$ and the final state $\rho(p)$. As $\rho$ is a pure state, the fidelity is given by
\begin{equation}
\begin{split}
F(\rho, \rho(t)) = & \braket{G | \rho(p) | G} = \bra{G} E_{0, 0} \rho E_{0, 0}^\dagger \ket{G} + \sum_{s = 1}^{N - 1} \bra{G} E_{0, s} \rho E_{0, s}^\dagger \ket{G} \\
= & (1 - p) + \frac{p}{N - 1} \sum_{s = 1}^{N - 1} \bra{G} U_{0, s} \rho U_{0 s}^\dagger \ket{G}.
\end{split}
\end{equation}
Using equation (\ref{Fidelity_half}) we have $F(\rho, \rho(t))$
\begin{equation}
= (1 - p) + \frac{p}{N^2(N - 1)} \sum_{s = 1}^{N - 1} \left| \sum_{i = 0}^{N - 1} (-1)^{g(i \oplus s) + g(i)} \right|^2,
\end{equation}
which depends on the function $g$. Below, we obtain an upper bound on the value of the fidelity.
\begin{equation}
F(\rho, \rho(t)) \leq (1 - p) + \frac{p}{N^2(N - 1)} N(N - 1) = 1 - p + \frac{p}{N} = 1 - \frac{p(N - 1)}{N}.
\end{equation}

\subsection{N-phase-flip noise} 
A qudit $\ket{i}$ under the influence of the $N$-phase-flip noise may be changed to any of the $(N - 1)$ possible states with probability $p$. The corresponding Kraus operators are of the form 
\begin{equation}
E_{r, 0} = \begin{cases} \sqrt{1 - p} I & ~\text{when}~ r = 0; \\ \sqrt{\frac{p}{N - 1}} U_{r, 0} & ~\text{for}~ 1 \leq r \leq (N - 1), s = 0. 
\end{cases}
\end{equation}
The new state after the application of the $N$-phase-flip noise is
\begin{equation}
\rho(p) = \sum_{r = 0}^{N - 1} E_{r, 0} \rho E_{r, 0}^\dagger = E_{0, 0} \rho E_{0, 0}^\dagger + \sum_{r = 1}^{N - 1} E_{r, 0} \rho E_{r, 0}^\dagger = (1 - p)\rho + \frac{p}{N - 1} \sum_{r= 1}^{N - 1} U_{r, 0} \rho U_{r, 0}^\dagger.
\end{equation}
Using equation (\ref{qudit_density_in_full_form}) along with equation (\ref{Weyl_on_G_density}), the state $\rho(p)$ is seen to be
\begin{equation}
\begin{split}
\rho(p) & = (1 - p) \sum_{i = 0}^{N  -1} \sum_{j = 0}^{N - 1} \frac{(-1)^{g(i) + g(j)}}{N} \ket{i}\bra{j} + \frac{p}{N - 1} \sum_{r = 1}^{N - 1} \left[\frac{1}{N} \sum_{i = 0}^{N - 1} \sum_{j = 0}^{N - 1} (-1)^{g(i) + g(j)} \omega_{N}^{(i - j)r} \ket{i}\bra{j} \right] \\
& = \sum_{i = 0}^{N  -1} \sum_{j = 0}^{N - 1} \frac{(-1)^{g(i) + g(j)}}{N} \left[ (1 - p) + \frac{p}{N(N - 1)} \sum_{r = 1}^{N - 1} \omega_{N}^{(i - j)r} \right] \ket{i}\bra{j}.
\end{split}
\end{equation}
As $\omega_{N} = \exp(\frac{2 \pi \iota}{N})$ is the primitive $N$-th root of unity we have $\sum_{r = 0}^{N - 1} \omega_{N}^{(i - j) r} = 0$, that is $\sum_{r = 1}^{N - 1} \omega_{N}^{(i - j) r} = -1$. Therefore, 
\begin{equation}
\rho(p) = \sum_{i = 0}^{N  -1} \sum_{j = 0}^{N - 1} \left[(1 - p) - \frac{p}{N - 1}\right]\frac{(-1)^{g(i) + g(j)}}{N} \ket{i}\bra{j}.
\end{equation}			 

The $(i,j)$-th term of $\rho(p)$ is represented by
\begin{equation}
\rho_{i,j} = \left[1 - p - \frac{p}{N - 1}\right]\frac{(-1)^{g(i) + g(j)}}{N}.
\end{equation}
Therefore the $l_1$ norm of coherence is
\begin{equation}
C_{l_1}(\rho(p)) = \frac{1}{N} \sum_{i = 0}^{N - 1} \sum_{\substack{j = 0 \\ j \neq i}}^{N - 1} \left|1 - p - \frac{p}{N - 1}\right| = \frac{N(N - 1)}{N} \left|1 - p - \frac{p}{N - 1}\right| = (N - 1)\left|1 - p - \frac{p}{N - 1}\right|. 
\end{equation} 
We can verify that $\left|1 - p - \frac{p}{N - 1}\right| < 1$. Thus, $C_{l_1}(\rho(p)) < (N - 1)$, which is the coherence of $\rho$. Hence, coherence decreases during the phase flip operation.

Fidelity between the states $\rho$ and $\rho(p)$ is
\begin{equation}
\begin{split}
F(\rho(p), \rho) & = \braket{G | E_{0, 0} |G} \braket{G | E_{0, 0}^\dagger|G} + \sum_{r = 1}^{N - 1} \braket{G| E_{r, 0} |G} \braket{G| E_{r, 0}^\dagger | G} \\
& = (1 - p) + \frac{p}{N - 1} \sum_{r = 1}^{N - 1} \bra{G} U_{r,0} \rho U_{r,0}^\dagger \ket{G}.
\end{split}
\end{equation}
From equation (\ref{Fidelity_half}) we see that
\begin{equation}
\begin{split}
\braket{G| U_{r, 0} \rho  U_{r, 0}^\dagger | G} & = \frac{1}{N^2} \left| \sum_{i = 0}^{N - 1} (-1)^{g(i) + g(i)} \omega_{N}^{ir} \right|^2 = \frac{1}{N^2} \left| \sum_{i = 0}^{N - 1} \omega_{N}^{ir} \right|^2 = 0,
\end{split}
\end{equation}
as $\sum_{i = 0}^{N - 1} \omega_{N}^{ir} = 0$. It indicates, $F(\rho(p), \rho) = (1 - p)$. Hence, fidelity is seen to depend on the noise parameter $p$. This could be attributed to the nature of the $N$-phase-flip noise which is an 
extension of the $\sigma_z$ operation, and hence acts equally on all states.

\subsection{Dit-phase-flip noise}

The dit-phase-flip noise is a combination of both the dit-flip and the phase-flip noises.
It is characterized by the Kraus operators 
\begin{equation}\label{dit-phase-flip noise}
E_{r, s} = \begin{cases} \sqrt{1 - p} I_{N} & ~\text{when}~ r = 0; \\ \sqrt{\frac{p}{N^2 - 1}} U_{r, s} & ~\text{for}~ 1 \leq r \leq (N - 1), s = 0. 
\end{cases}
\end{equation}
As a result, after performing the dit-flip operation and using Eq. \eqref{Kraus_operator},(\ref{qudit_density_in_full_form}) and (\ref{Weyl_on_G_density}) the new state will be

\begin{equation}
\rho(p) = \frac{1}{N}\sum_{i = 0}^{N  -1} \sum_{j = 0}^{N - 1} \left[(1 - p) (-1)^{g(i) + g(j)} + \frac{p}{N^2 - 1} \sum_{r = 0}^{N - 1} \sum_{\substack{s = 0 \\ (r, s) \neq (0, 0)}}^{N - 1} (-1)^{g(i \oplus s) + g(j \oplus s)} \omega_{N}^{(i - j) r} \right] \ket{i}\bra{j}.
\end{equation}
Now, we calculate the $l_1$ norm of coherence of the new state $\rho(p)$. The $(i,j)$-th entry of $\rho(p)$ is 
\begin{equation}
\rho_{i,j} = \frac{1}{N}\left[(1 - p) (-1)^{g(i) + g(j)} + \frac{p}{N^2 - 1} \sum_{r = 0}^{N - 1} \sum_{\substack{s = 0 \\ (r, s) \neq (0, 0)}}^{N - 1} (-1)^{g(i \oplus s) + g(j \oplus s)} \omega_{N}^{(i - j) r} \right].
\end{equation}
The absolute value is bounded by
\begin{equation}
|\rho_{i,j}| \leq \frac{1}{N}\left[(1 - p) + \frac{p}{N^2 - 1} \sum_{r = 0}^{N - 1} \sum_{\substack{s = 0 \\ (r, s) \neq (0, 0)}}^{N - 1} 1 \right] = \frac{1}{N}\left[(1 - p) + \frac{p}{N^2 - 1} (N^2 - 1) \right] = \frac{1}{N}.
\end{equation}

Therefore the $l_1$ norm of coherence of $\rho(p)$ will be bounded by
\begin{equation}
C_{l_1}(\rho(p)) = \sum_{i = 0}^{N - 1} \sum_{\substack{j = 0 \\ j \neq i}}^{N - 1} |\rho_{i,j}| \leq \frac{1}{N} \times N(N - 1) = N - 1,
\end{equation}
which is the $l_1$ norm of coherence of the original state. Thus, coherence decreases under the application of the dit-phase-flip noise.

Now, we calculate the fidelity between the initial state and the resultant state after applying the dit-phase-flip noise. 
\begin{equation}
\begin{split}
F(\rho(p), \rho)
& = (1 - p)+ \frac{p}{N^2(N^2 - 1)} \sum_{r = 0}^{N - 1} \sum_{\substack{s = 0 \\ (r, s) \neq (0, 0)}}^{N - 1} \left| \sum_{i = 0}^{N - 1} (-1)^{g(i \oplus s) + g(i)} \omega_{N}^{ir} \right|^2,
\end{split}
\end{equation}  
So the fidelity $F(\rho(p), \rho)$ depends on the function $g$, in the particular state under consideration.

\subsection{Depolarizing noise}

The Kraus operators generating a depolarizing channel is represented by \cite{imany2019high, gokhale2019asymptotic} 
\begin{equation}\label{Depolarizing noise}
E_{r, s} = \begin{cases} \sqrt{1 - \frac{N^2 - 1}{N^2}p} I_{N} & ~\text{when}~ r = 0, s = 0; \\ \frac{\sqrt{p}}{N} U_{r, s} & ~\text{for}~ 0 \leq r, s \leq (N - 1) ~\text{and}~ (r, s) \neq (0, 0). \end{cases}
\end{equation}
Expanding $\rho$ and $U_{r, s} \rho U_{r, s}^\dagger$ using equations (\ref{qudit_density_in_full_form}) and (\ref{Weyl_on_G_density}), respectively, the new state after application of the depolarizing noise is

\begin{equation}
\begin{split}
\rho(p)
& = \sum_{i = 0}^{N  -1} \sum_{j = 0}^{N - 1} \left[ \left(1 - \frac{N^2 - 1}{N^2}p \right) \frac{(-1)^{g(i) + g(j)}}{N}  + \frac{p}{N^3} \sum_{r = 0}^{N - 1} \sum_{\substack{s = 0 \\ (r, s) \neq (0, 0)} }^{N - 1} (-1)^{g(i \oplus s) + g(j \oplus s)} \omega_{N}^{(i - j) r}  \right] \ket{i}\bra{j}.
\end{split}
\end{equation}
Now we work out the $l_1$ norm of coherence in the state $\rho(p)$. The $(i,j)$-th element of $\rho(p)$ is given by
\begin{equation}
\rho_{i,j} = \left(1 - \frac{N^2 - 1}{N^2}p \right) \frac{(-1)^{g(i) + g(j)}}{N} + \frac{p}{N^3} \sum_{r = 0}^{N - 1} \sum_{\substack{s = 0 \\ (r, s) \neq (0, 0)} }^{N - 1} (-1)^{g(i \oplus s) + g(j \oplus s)} \omega_{N}^{(i - j) r}.
\end{equation}
An upper bound on the absolute values of $\rho_{i,j}$ is given by

\begin{equation} 
|\rho_{i,j}| \leq \left(1 - \frac{N^2 - 1}{N^2}p \right) \frac{1}{N} + \frac{p(N^2 - 1)}{N^3} = \frac{1}{N}.
\end{equation}
Therefore the $l_1$ norm of coherence is bounded by
\begin{equation}
C_{l_1}(\rho(p)) = \sum_{i = 0}^{N - 1} \sum_{\substack{j = 0 \\ j \neq i}}^{N - 1} |\rho_{i,j}| \leq \frac{N (N - 1)}{N} = N - 1,
\end{equation}
which is the coherence of the original state. Therefore, coherence decreases when we apply the depolarizing operation on the states. 
Applying equations (\ref{Fidelity_half}) and (\ref{Fidelity}), the fidelity between the state $\rho(p)$ and $\rho$ is seen to be
\begin{equation}
\begin{split}
F(\rho(p), \rho)
= & \left(1 - \frac{N^2 - 1}{N^2}p \right) + \frac{p}{N^4} \sum_{r = 0}^{N - 1} \sum_{\substack{s = 0 \\ (r, s) \neq (0, 0)} }^{N - 1} \left| \sum_{i = 0}^{N - 1} (-1)^{g(i \oplus s) + g(i)} \omega_{N}^{ir} \right|^2.
\end{split}
\end{equation}
Therefore, $F(\rho(p), \rho)$ depends on the number of vertices, and the structure of $G$, as well as the channel parameter $p$.	

\subsection{Amplitude Damping Channel (non-Markovian)} 

\begin{figure}
	\centering
	\includegraphics[scale = .6]{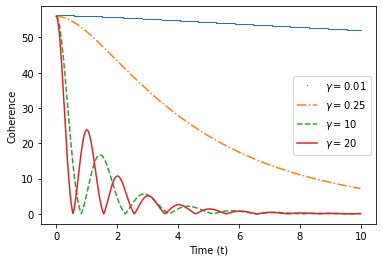}
	\caption{(Color online) Coherence of the state generated by Amplitude Damping Channel (non-Markovian) as a function of $t$. Here $g = 1$ in all the cases. In the Markovian domain, we consider $\gamma = 0.01$ and $\gamma = 0.25$. In non-Markovian domain we consider $\gamma = 10$ and $\gamma = 20$.}
	\label{coherence_plot} 
\end{figure}

The non-Markovian Amplitude Damping Channel (ADC) for qubits is characterized by the Kraus operators \cite{ghosal2021characterizing, utagi2020ping}  
\begin{equation}
\begin{split} 
& M_0 = \begin{bmatrix} 1 & 0 \\ 0 & \sqrt{1 - \lambda(t)} \end{bmatrix} ~\text{and}~ M_1 = \begin{bmatrix} 0 & \sqrt{\lambda(t)} \\ 0 & 0 \end{bmatrix}; \\
\end{split} 
\end{equation}
where $\lambda(t) = 1 - e^{-gt}\left(\frac{g}{l} \sinh \left[\frac{lt}{2}\right] + \cosh \left[\frac{lt}{2}\right]\right)^2$, and $l = \sqrt{g^2 - 2 \gamma g}$. The system exhibits Markovian and non-Markovian evolution of a state when $2 \gamma \ll g$ and $2 \gamma \gg g$, respectively, in the case of qubit states. It can be easily seen that
\begin{equation}
\sqrt{1 - \lambda(t)} = e^{-\frac{gt}{2}} \left[\frac{g}{l}\sinh \left(\frac{lt}{2}\right) + \cosh \left(\frac{lt}{2}\right)\right].
\end{equation}
Note that, $\sqrt{1 - \lambda(t)} > 0$.
For an $N$-dit system we generalize the ADC non-Markovian channel using the Kraus operators 
\begin{equation}
\begin{split}
E_0 & = \ket{0}\bra{0} + \sqrt{1 - \lambda(t)} \sum_{i = 1}^{N - 1} \ket{i} \bra{i} \\
E_i & = \sqrt{\lambda(t)} \ket{0}\bra{i} ~\text{for}~ 1 \leq i \leq N - 1.
\end{split}
\end{equation}
These satisfy $E_0^\dagger E_0 = \ket{0}\bra{0} + \left(1 - \lambda(t) \right) \sum_{i = 1}^{N - 1} \ket{i} \bra{i}$ and $E_i^\dagger E_i = \lambda(t) \ket{i}\bra{i}$ for $i = 1, 2, 3, \dots (N - 1)$. Therefore, 
\begin{equation}
\sum_{i = 0}^{N - 1} E_i^\dagger E_i = \ket{0}\bra{0} + \left(1 - \lambda(t) \right) \sum_{i = 1}^{N - 1} \ket{i} \bra{i} + \lambda(t) \sum_{i = 1}^{N - 1}  \ket{i}\bra{i} = \ket{0}\bra{0} + \sum_{i = 1}^{N - 1} \ket{i} \bra{i} = I_{N}.
\end{equation} 
This indicates that $E_i$ are bonafide Kraus operators. Next, we apply these Kraus operators on the state $\ket{G}$, equation (\ref{qudit_hypergraph_state}). Note that, $E_0 \ket{G} = \frac{1}{\sqrt{N}} \left[ \ket{0} + \sqrt{1 - \lambda(t)} \sum_{i = 1}^{N - 1} (-1)^{g(i)} \ket{i} \right]$. Therefore,
\begin{equation}
\begin{split} 
E_0 \rho E_0^\dagger = & E_0 \ket{G} \bra{G} E_0^t = \frac{1}{N} [ \ket{0} \bra{0} + \sqrt{1 - \lambda(t)} \sum_{j = 1}^{N - 1} (-1)^{g(j)} \ket{0}\bra{j} \\
& + \sqrt{1 - \lambda(t)} \sum_{i = 1}^{N - 1} (-1)^{g(i)} \ket{i}\bra{0} + (1 - \lambda(t)) \sum_{i = 1}^{N - 1} \sum_{j = 1}^{N - 1} (-1)^{g(i) + g(j)} \ket{i} \bra{j} ].
\end{split} 
\end{equation} 
Also, for $1 \leq i \leq N - 1$, we have $E_i \ket{G} = \frac{\sqrt{\lambda(t)}}{\sqrt{N}} (-1)^{g(i)}\ket{0}$. Hence, $E_i \rho E_i^\dagger = E_i \ket{G}\bra{G} E_i^\dagger = \frac{\lambda(t)}{N} \ket{0}\bra{0}$ for $i = 1, 2, \dots (N - 1)$. Combining we get,
\begin{equation}
\begin{split}
\rho(t) & = \sum_{i = 0}^{N - 1} E_i \rho E_i^t = \frac{1}{N} [ \ket{0} \bra{0} + \sqrt{1 - \lambda(t)} \sum_{j = 1}^{N - 1} (-1)^{g(j)} \ket{0}\bra{j} + \sqrt{1 - \lambda(t)} \sum_{i = 1}^{N - 1} (-1)^{g(i)} \ket{i}\bra{0}\\
& \hspace{4cm} + (1 - \lambda(t)) \sum_{i = 1}^{N - 1} \sum_{j = 1}^{N - 1} (-1)^{g(i) + g(j)} \ket{i} \bra{j} ] + (N - 1) \frac{\lambda(t)}{N} \ket{0}\bra{0}\\
& = \left[(N - 1) \frac{\lambda(t)}{N} + \frac{1}{N} \right] \ket{0}\bra{0} + \sqrt{1 - \lambda(t)} \sum_{j = 1}^{N - 1} (-1)^{g(j)} \ket{0}\bra{j} + \sqrt{1 - \lambda(t)} \sum_{i = 1}^{N - 1} (-1)^{g(i)} \ket{i}\bra{0}\\
& \hspace{4cm} + (1 - \lambda(t)) \sum_{i = 1}^{N - 1} \sum_{j = 1}^{N - 1} (-1)^{g(i) + g(j)} \ket{i} \bra{j}.
\end{split}
\end{equation} 
The expression of $\rho(t)$ indicates that $(i,j)$-th term of $\rho(t)$ for $i \neq j$ is represented by
\begin{equation}
\rho_{i,j} = \begin{cases} (-1)^{g(j)} \sqrt{1 - \lambda(t)} & ~\text{when}~ i = 0, ~\text{and}~ j = 1, 2, \dots (N - 1); \\ 
(-1)^{g(i)} \sqrt{1 - \lambda(t)} & ~\text{when}~ j = 0, ~\text{and}~ i = 1, 2, \dots (N - 1); \\ 
(-1)^{g(i) + g(j)} (1 - \lambda(t)) & ~\text{when}~ i \neq 0, j\neq 0, i\neq j, ~\text{and}~ i, j = 1, 2, \dots (N - 1). \end{cases} 
\end{equation}
The $l_1$ norm of coherence of the state $C_{l_1}(\rho(t))$ is
\begin{equation}
C_{l_1}(\rho(t)) = \sum_{i = 0}^{N - 1} \sum_{\substack{j = 0\\ j \neq i}}^{N - 1} |\rho_{i,j}| = (N - 1)\left[ 2\sqrt{1 - \lambda(t)} + (N - 2) (1 - \lambda(t)) \right].
\end{equation} 
Note that, the expression of $C_{l_1}(\rho(t))$ depends only on $N$ and $\lambda(t)$ coming from the noisy channel. Recall from equation (\ref{coherence_of_hypergraph_state}) that the coherence of $\rho$ is $(N - 1)$. Coherence decreases under the ADC noise if $2\sqrt{1 - \lambda(t)} + (N - 2) (1 - \lambda(t)) < 1$. Simplifying we get
\begin{equation}
\frac{-1 - \sqrt{N - 1}}{N - 2} < \sqrt{1 - \lambda(t)} < \frac{-1 + \sqrt{N - 1}}{N - 2}.
\end{equation}
Therefore,the coherence of $\rho(t)$ is less than the coherence of original state when $\sqrt{1 - \lambda} < \frac{-1 + \sqrt{N - 1}}{N - 2}$. For $n = 3$, we have plotted
the coherence in figure \ref{coherence_plot}. In the non-Markovian regime, we can see the typical recurrence behavior, due to the $P$-indivisible nature of the noise \cite{utagi2020temporal}.

\begin{figure}
	\centering 
	\includegraphics[scale = .6]{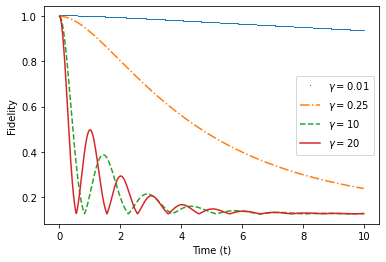}
	\caption{(Color online) Fidelity between the states and the state generated by Amplitude Damping Channel (non-Markovian) as a function of $t$. Here, we consider $g = 1$ in all the cases. In the Markovian domain, we consider $\gamma = 0.01$ and $\gamma = 0.25$. In non-Markovian domain we consider $\gamma = 10$ and $\gamma = 20$.}
	\label{fidelity_plot} 
\end{figure}

The fidelity between $\rho(t)$ and $\rho$ is given by
\begin{equation}
F(\rho(t), \rho) = \braket{G | \rho(t) | G} = \braket{G | E_0 | G } \braket{G | E_0^{\dagger} | G} + \sum_{i = 1}^{N - 1} \braket{G | E_i | G} \braket{G | E_i^{\dagger}
| G}.
\end{equation} 
As $E_0 \ket{G} = \frac{1}{\sqrt{N}} \left[ \ket{0} + \sqrt{1 - \lambda(t)} \sum_{i = 1}^{N - 1} (-1)^{g(i)} \ket{i} \right]$, we have 
\begin{equation}
\begin{split}
\braket{G | E_0 | G} = \frac{1}{\sqrt{N}} \left[ \frac{1}{\sqrt{N}} + \sqrt{1 - \lambda(t)} \sum_{i = 1}^{N - 1} \frac{(-1)^{g(i)} (-1)^{g(i)}}{\sqrt{N}} \right] = \frac{1}{N} \left[ 1 + (N - 1) \sqrt{1 - \lambda(t)} \right].
\end{split}
\end{equation}
Also, $E_i \ket{G} = \frac{\sqrt{\lambda(t)}}{\sqrt{N}} (-1)^{g(i)}\ket{0}$ for $1 \leq i \leq N - 1$. Hence, $\braket{G | E_i | G} = \frac{\sqrt{\lambda(t)}}{N} (-1)^{g(i)}$. Combining we get 
\begin{equation}
F(\rho(t), \rho) = \frac{1}{N^2} \left[ 1 + (N - 1) \sqrt{1 - \lambda(t)} \right]^2 + \frac{\lambda(t)}{N^2} (N - 1).
\end{equation}
In this case, fidelity depends on the noise characterized by $\lambda(t)$. It does not depend 
on the structural properties of $G$. For 
different values of $n$ the fidelity is shown in figure \ref{fidelity_plot}. Recurrences, due to the $P$-indivisible nature 
of the noise, are seen in the non-Markovian regime.

\subsection{Non-Markovian Dephasing} 
For a qubit system the non-Markovian dephasing channel has been studied in \cite{shrikant2018non}. We generalize the non-Markovian dephasing channel for a qudit system using the following Kraus operators,
\begin{equation}\label{Non-Markovian dephasing}
E_{r, s} = \begin{cases} \sqrt{1 - \kappa } I & ~\text{when}~ r = 0, s = 0; \\ \sqrt{\frac{\kappa }{N^2 - 1}} U_{r, s} & ~\text{for}~ 0 \leq r, s \leq (N - 1) ~\text{and}~ (r, s) \neq (0, 0). \end{cases}
\end{equation}
where we have
$\kappa = [ 1 + \alpha(1 - p)] p,$ $0 \leq p \leq \frac{1}{2}$ and $0 \leq \alpha \leq 1$. The non-Markovianity of the channel depends on the choice of the value of $\alpha$ and the function $\kappa$ of $p$. Here, we consider 
\begin{equation}
\kappa(p) = p \frac{1 + \eta (1 - 2p) \sin(\omega p)}{1 + \eta (1 - 2p)}.
\end{equation}
Here $\eta$ and $\omega$ are two positive constants characterizing the strength and frequency of the channel. Also, $0 \leq p \leq \frac{1}{2}$. 

Clearly, $E_{0, 0}^\dagger E_{0, 0} = (1 - \kappa) I$, and $E_{r, s}^\dagger E_{r, s} = \frac{\kappa }{N^2 - 1} U_{r, s}^\dagger U_{r, s} = \frac{\kappa }{N^2 - 1}I$. Combining we get
\begin{equation}
\sum_{r = 0}^{N - 1} \sum_{s = 0}^{N - 1} E_{r, s}^\dagger E_{r, s} = (1 - \kappa) I + \frac{\kappa }{N^2 - 1} (N^2 - 1)I = I,
\end{equation}
justifying that $E_{r,s}$ are Kraus operators.

Now we apply these Kraus operators on the state $\rho = \ket{G}\bra{G}$. The new state is
\begin{equation}
\rho(\kappa) = \sum_{r = 0}^{N - 1} \sum_{s = 0}^{N - 1} E_{r, s} \rho E_{r, s}^\dagger
= \frac{1}{N} \sum_{i = 0}^{N  -1} \sum_{j = 0}^{N - 1} \left[ (-1)^{g(i) + g(j)} (1 - \kappa) + \frac{\kappa }{N^2 - 1} \sum_{r = 0}^{N - 1} \sum_{\substack{s = 0 \\ (r, s) \neq (0, 0)}}^{N - 1} (-1)^{g(i \oplus s) + g(j \oplus s)} \omega_{N}^{(i - j)r} \right] \ket{i}\bra{j},
\end{equation}

obtained by an application of equations (\ref{qudit_density_in_full_form}) and (\ref{Weyl_on_G_density}).

Next, we study the coherence of the state $\rho(\kappa)$. The $(i,j)$-th term of the $\rho(\kappa)$ is 
\begin{equation}
\rho_{i,j} = \frac{1}{N} [ (-1)^{g(i) + g(j)} (1 - \kappa) + \frac{\kappa }{N^2 - 1} \sum_{r = 0}^{N - 1} \sum_{\substack{s = 0 \\ (r, s) \neq (0, 0)}}^{N - 1} (-1)^{g(i \oplus s) + g(j \oplus s)} \omega_{N}^{(i - j)r} ].
\end{equation}
The absolute value $|\rho_{i,j}|$ is bounded by 
\begin{equation}
|\rho_{i,j}| \leq \frac{(1 - \kappa)}{N} + \frac{\kappa }{N(N^2 - 1)} \sum_{r = 0}^{N - 1} \sum_{\substack{s = 0 \\ (r, s) \neq (0, 0)}}^{N - 1} 1 = \frac{(1 - \kappa)}{N} + \frac{\kappa }{N(N^2 - 1)} (N^2 - 1) = \frac{1}{N}.
\end{equation}
Therefore, the $l_1$ measure of coherence is bounded by
\begin{equation}
C_{l_1}(\rho(\kappa)) = \sum_{i = 0}^{N - 1} \sum_{\substack{j = 0 \\ j \neq i}}^{N - 1} |\rho_{i,j}| \leq \frac{(N^2 - N)}{N} = N - 1,
\end{equation}
which is the coherence of the state $\rho$. Thus, coherence decreases under non-Markovian Dephasing operations.

\begin{figure}
	\centering 
	\includegraphics[scale = .45]{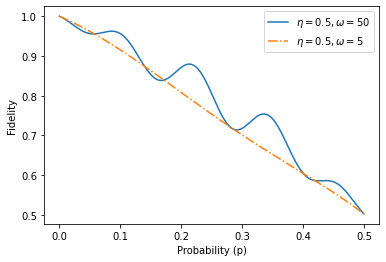}
	\caption{(Color online) Fidelity between the hypergraph states and the final state generated by the non-Markovian Dephasing channel is plotted 
	as a function of $p$ for the hypergraph depicted in \ref{example_hypergraph}. In the Markovian case we consider $\omega = 5$ and in 
	the non-Markovian case $\omega = 50$.}
	\label{fidelity_non_markovian_dephasing} 
\end{figure}

Fidelity between the state $\rho(\kappa)$ and $\rho$ is represented by
\begin{equation}
\begin{split}
F(\rho(\kappa), \rho)
= & (1 - \kappa) + \frac{\kappa }{N^2(N^2 - 1)} \sum_{r = 0}^{N - 1} \sum_{\substack{s = 0 \\ (r, s) \neq (0, 0)}}^{N - 1} \left| \sum_{i = 0}^{N - 1} (-1)^{g(i \oplus s) + g(i)} \omega_{N}^{ir} \right|^2,
\end{split}
\end{equation}
applying equation (\ref{Fidelity_half}). Clearly, $F(\rho(\kappa), \rho)$ depends on the structure of $G$ and the number of vertices in it. Also, it depends on the 
channel parameter $\kappa$. The fidelity in Markovian and non-Markovian domains is depicted in figure \ref{fidelity_non_markovian_dephasing}.
The characteristic recurrent behavior in the non-Markovian regime reflects upon the $P$-indivisible nature of the noise here.
\subsection{Non-Markovian depolarization}  
We define the non-Markovian depolarization noise with the Kraus operators \cite{shrikant2018non}:
\begin{equation}\label{non-Markovian depolarization}
E_{r, s} = \begin{cases} \sqrt{1 + \frac{(N^2 - 1)(1 - p)}{N^2}\Lambda_1} I_N & ~\text{when}~ r = 0, s = 0; \\ \frac{\sqrt{p \Lambda_2}}{N} U_{r, s} & ~\text{for}~ 0 \leq r, s \leq (N - 1), ~\text{and}~ (r, s) \neq (0, 0),\end{cases}
\end{equation}

where $\Lambda_1 = -\alpha p$ and $\Lambda_2 = \alpha (1 - p)$. Note that $(1 - p)\Lambda_1 + p \Lambda_2 = 0$. Note that 
\begin{equation}
\begin{split}
\sum_{r = 0}^{N - 1} \sum_{s = 0}^{N - 1} E_{r, s}^\dagger E_{r, s} & = E_{0, 0}^\dagger E_{0, 0} + \sum_{r = 0}^{N - 1} \sum_{\substack{s = 0 \\ (r,s) \neq (0, 0)}}^{N - 1} E_{r, s}^\dagger E_{r, s} = \left(1 + \frac{(N^2 - 1)(1 - p)}{N^2}\Lambda_1 \right) I_N + \sum_{r = 0}^{N - 1} \sum_{\substack{s = 0 \\ (r,s) \neq (0, 0)}}^{N - 1} \frac{p \Lambda_2}{N^2} U_{r, s}^\dagger U_{r, s} \\
& = \left(1 + \frac{(N^2 - 1)(1 - p)}{N^2}\Lambda_1 \right) I_N + \frac{(N^2 - 1)p \Lambda_2}{N^2} I_N = I_N + \frac{N^2 - 1}{N^2} [(1 - p)\Lambda_1 + p \Lambda_2] = I_N.
\end{split}
\end{equation}
It justifies $E_{r,s}$ as bonafide Kraus operators.

Applying these Kraus operators on $\ket{G}$ along with equations (\ref{qudit_density_in_full_form}) and (\ref{Weyl_on_G_density}), we have the new state using equation (29),
\begin{equation}
\begin{split}
\rho(\alpha)=\frac{1}{N} \sum_{i = 0}^{N  -1} \sum_{j = 0}^{N - 1} \left[ (-1)^{g(i) + g(j)} \left(1 + \frac{(N^2 - 1)(1 - p)}{N^2}\Lambda_1 \right) + \frac{p \Lambda_2}{N^2} \sum_{r = 0}^{N - 1} \sum_{\substack{s = 0 \\ (r,s) \neq (0, 0)}}^{N - 1} (-1)^{g(i \oplus s) + g(j \oplus s)} \omega_{N}^{(i - j)r} \right] \ket{i}\bra{j}.
\end{split}
\end{equation}

Now we calculate the coherence of the state $\rho(\alpha)$. The $(i,j)$-th element of $\rho(\alpha)$ is 
\begin{equation}
\rho_{i,j} = \frac{1}{N} [(-1)^{g(i) + g(j)} \left(1 + \frac{(N^2 - 1)(1 - p)}{N^2}\Lambda_1 \right) + \frac{p \Lambda_2}{N^2} \sum_{r = 0}^{N - 1} \sum_{\substack{s = 0 \\ (r,s) \neq (0, 0)}}^{N - 1} (-1)^{g(i \oplus s) + g(j \oplus s)} \omega_{N}^{(i - j)r} ].
\end{equation}
{The absolute value of $\rho_{i,j}$ is bounded by
\begin{equation}
\begin{split}
|\rho_{i,j}| & \leq \frac{1}{N} \left(1 + \frac{(N^2 - 1)(1 - p)}{N^2}\Lambda_1 \right) + \frac{p \Lambda_2}{N^3} (N^2 - 1)  = \frac{1}{N} + \frac{(N^2 - 1)}{N^3}\left[ (1 - p)\Lambda_1 + p \Lambda_2 \right] = \frac{1}{N}.
\end{split}
\end{equation}
Therefore, the $l_1$ norm of coherence is bounded by
\begin{equation}
C_{l_1}(\rho(\alpha)) = \sum_{i = 0}^{N - 1} \sum_{\substack{j = 0 \\ j \neq i}}^{N - 1} |\rho_{i, j}| \leq \frac{N^2 - N}{N} = (N - 1),
\end{equation}
which is the coherence of the state $\rho$. Therefore, coherence decreases under the non-Markovian depolarization noise. 

The fidelity between the states $\rho$ and $\rho(\alpha)$ is seen to be after applying equations (\ref{Fidelity_half}) and (\ref{Fidelity}),
\begin{equation}
\begin{split}
F(\rho, \rho(\alpha)) & = \braket{G | \rho(\alpha) |G }= \left(1 + \frac{(N^2 - 1)(1 - p)}{N^2}\Lambda_1 \right) + \frac{p \Lambda_2}{N^4} \sum_{r = 0}^{N - 1} \sum_{\substack{s = 0 \\ (r,s) \neq (0, 0)}}^{N - 1} \left| \sum_{i = 0}^{N - 1} (-1)^{g(i \oplus s) + g(i)} \omega_{N}^{ir} \right|^2,
\end{split}
\end{equation}
Clearly, $F(\rho, \rho(\alpha))$ depends on $N$, as well as the channel parameters $\Lambda_1$ and $\Lambda_2$.

\subsection{Noisy channels in generalized form}

In general we can write the Kraus operators as
	\begin{equation}\label{generalised_Kraus_operators}
		E_{r, s} = \begin{cases} \alpha I_{N} & ~\text{when}~ r = 0, s = 0; \\ \beta U_{r, s} & ~\text{for}~ 0 \leq r, s \leq (N - 1) ~\text{and}~ (r, s) \neq (0, 0). \end{cases}
	\end{equation}
	Different values of $\alpha$ and $\beta$ in equation (\ref{generalised_Kraus_operators}) generates different Kraus operators if the following condition (the completeness relation) is satisfied.
     \begin{equation}
     \sum_{r = 0}^{N - 1} \sum_{s = 0}^{N - 1} E_{r, s}^\dagger E_{r, s}=I_N
      \end{equation}
      we have \begin{equation}
          E_{0, 0}^\dagger E_{0, 0} + \sum_{r = 0}^{N - 1} \sum_{\substack{s = 0 \\ (r,s) \neq (0, 0)}}^{N - 1} E_{r, s}^\dagger E_{r, s} = \alpha^\dagger\alpha I_N + \sum_{r = 0}^{N - 1} \sum_{\substack{s = 0 \\ (r,s) \neq (0, 0)}}^{N-1}\beta^\dagger\beta U_{r, s}^\dagger U_{r, s} 
          =[|\alpha|^2 + |\beta|^2 \times (N^2-1)]I_N
      \end{equation}
      So equation (68) is satisfied if
      \begin{equation}\label{Completeness relation}
          |\alpha|^2 + |\beta|^2 \times (N^2-1) = 1
      \end{equation} 
      \\
     For instance, $\alpha = \sqrt{1 - p}$ and $\beta=\sqrt{\frac{p}{N^2 - 1}}$ returns dit-phase-flip noise mentioned in equation (\ref{dit-phase-flip noise}). Also, $\alpha = \sqrt{1 - \frac{N^2 - 1}{N^2}p}$ and $\beta =  \sqrt{\frac{p}{N}}$ generates the depolarizing noise mentioned in equations (\ref{Depolarizing noise}).. Similarly , comparing equations (\ref{Non-Markovian dephasing}) and (\ref{generalised_Kraus_operators}) we observe that the Non-Markovian dephasing noise is developed for $\alpha = \sqrt{1 - \kappa }$ and $\beta = \sqrt{\frac{\kappa }{N^2 - 1}}$. In addition, $\alpha = \sqrt{1 + \frac{(N^2 - 1)(1 - p)}{N^2}\Lambda_1}$ and $\beta =  \frac{\sqrt{p \Lambda_2}}{N}$ yield non-Markovian depolarization noise as discussed in equations (\ref{non-Markovian depolarization}).
     Here the parameter p associated with each the Kraus operator corresponding to any noise represents the probability of that specific noise process occuring. It quantifies the strength of that corresponding error. The value of p may vary, depending on the noise being considered. Its value ranges from 0 to 1, where p=0 indicates the absence of that particular noise process or error and p=1 indicates that the specific noise always occurs with certainty. By adjusting the values of the parameter p, we can control the relative stength of different noises.

	Applying the Kraus operators on $\ket{G}$, we get 
	\begin{equation}
		\rho(\alpha,\beta) =\sum_{r = 0}^{N - 1} \sum_{s = 0}^{N - 1} E_{r, s} \rho E_{r, s}^\dagger= \frac{1}{N}\sum_{i = 0}^{N  -1} 	\sum_{j = 0}^{N - 1} \left[\alpha^2(-1)^{g(i) + g(j)} + \beta^2\sum_{r = 0}^{N - 1} \sum_{\substack{s = 0 \\ (r, s) \neq (0, 0)}}^{N - 1} (-1)^{g(i \oplus s) + g(j \oplus s)} \omega_{N}^{(i - j) r} \right] \ket{i}\bra{j}.
	\end{equation}
	Now we calculate the $l_1$ norm of coherence of the new state $\rho(\alpha,\beta)$. The $(i,j)$-th entry of $\rho(\alpha,\beta)$ is 
	\begin{equation}
		\rho_{i,j} = \frac{1}{N}\left[\alpha^2(-1)^{g(i) + g(j)} + \beta^2\sum_{r = 0}^{N - 1} \sum_{\substack{s = 0 \\ (r, s) \neq (0, 0)}}^{N - 1} (-1)^{g(i \oplus s) + g(j \oplus s)} \omega_{N}^{(i - j) r} \right].
	\end{equation}
	The absolute value of $\rho_{i,j}$ is bounded by [using equation(\ref{Completeness relation}]
	\begin{equation}\label{rho_ij}
		|\rho_{i,j}| \leq \frac{1}{N}\left[\alpha^2+ \beta^2\sum_{r = 0}^{N - 1} \sum_{\substack{s = 0 \\ (r, s) \neq (0, 0)}}^{N - 1} 1 \right] = \frac{1}{N}\left[\alpha^2+ \beta^2(N^2 - 1) \right]=\frac{1}{N}
	\end{equation}
	Therefore the $l_1$ norm of coherence of $\rho(p)$ will be bounded by using equations (\ref{Completeness relation}) and (\ref{rho_ij}) 
	\begin{equation}
		C_{l_1}(\rho(\alpha,\beta)) = \sum_{i = 0}^{N - 1} \sum_{\substack{j = 0 \\ j \neq i}}^{N - 1} |\rho_{i,j}| \leq \frac{1}{N}\left[\alpha^2+ \beta^2(N^2 - 1) \right] \times N(N - 1) = \left[\alpha^2+ \beta^2(N^2 - 1) \right] \times (N-1)=(N-1)
	\end{equation}
which is the same for all four noises as described above in this section. Therefore the coherence decreases under the application of the noises.
	
	Similarly, the equation for the fidelity between the initial state and the final state is given by
	\begin{equation}\label{general_fidelity}
		\begin{split}
			F(\rho(\alpha,\beta), \rho) & = \braket{G | \rho(p) | G} = \alpha^2\braket{G | \rho | G} + \frac{\beta^2}{N^2} \sum_{r = 0}^{N - 1} \sum_{\substack{s = 0 \\ (r, s) \neq (0, 0)}}^{N - 1} \braket{G | U_{r, s} \rho U_{r, s}^\dagger | G}\\
			& = \alpha^2+ \frac{\beta^2}{N^2} \sum_{r = 0}^{N - 1} \sum_{\substack{s = 0 \\ (r, s) \neq (0, 0)}}^{N - 1} \left| \sum_{i = 0}^{N - 1} (-1)^{g(i \oplus s) + g(i)} \omega_{N}^{ir} \right|^2.
		\end{split}
	\end{equation} 

\section{Conclusion}
Due to their important applications in quantum computation, in 
particular, the intrinsic 
physical characteristics of the qudit states are worth investigating. In this work, we studied the effect of noise on qudit states. We applied a number of 
noisy quantum channels, both Markovian and non-Markovian, such as the dit-flip noise, phase flip noise, dit-phase flip noise, depolarizing noise, non-Markovian 
Amplitude Damping Channel (ADC), dephasing noise, and depolarization noise. We worked out the analytic expression of the final state after applying the noisy channel. 
In addition, we studied the change in coherence, as well as the fidelity between the initial and final state. The coherence decreases under the application of 
all these channels, except the non-Markovian ADC channel, where the phenomena of recurrences in the non-Markovian regime are responsible for this behavior. In 
case of non-Markovian ADC, coherence decreases if the channel parameter $\lambda$ satisfies the following inequality
\begin{equation}
\sqrt{1 - \lambda(t)} < \frac{-1 + \sqrt{N - 1}}{N - 2}, 
\end{equation}
where $N = 2^n$. 

	Quantum teleportation plays a pivotal role in the field of quantum information technology. It has numerous
	applications in quantum computing and quantum cryptography. Using qudits, the extension of original teleportation protocols has been performed. Continuous variable teleportation has also been realized in optical systems. Our results would be of use in this context, for example, in understanding the role of noise in quantum optical experiments pertaining to quantum teleportation.

\section*{Appendix: Analogy with quantum hypergraph states}

In combinatorics a simple graph $G = (V(G), E(G))$ is a combinatorial object consisting of a set of vertices $V(G)$, and edges $E(G)$. An edge in a graph is a set of 
two vertices. A hyperedge is a set composed of more than two vertices. A hypergraph \cite{bretto2013hypergraph} is a generalization of a graph and is a combination of 
a set of vertices $V(H)$ and a set of hyperedges $E(H)$, which is denoted by $H = (V(H), E(H))$. An example of a hypergraph is depicted in 
figure \ref{example_hypergraph}.
\begin{figure}
	\centering
	\begin{subfigure}{0.4\textwidth}
		\centering 
		\begin{tikzpicture}[scale = 2.5]
		\draw[fill] (0, 1) circle [radius = 1.2pt];
		\node[below right] at (0, 1) {$0$};
		\draw[fill] (1, 1) circle [radius = 1.2pt];
		\node[below left] at (1, 1) {$1$};
		\draw[fill] (1, 0) circle [radius = 1.2pt];
		\node[above left] at (1, 0) {$2$};
		\draw[fill] (0, 0) circle [radius = 1.2pt];
		\node[above right] at (0, 0) {$3$};
		\draw (0, 1)--(0,0);
		\draw [rounded corners] (-0.1, .5) -- (-0.1, 1.1) -- (-.1, -.1) -- (1.25, -.1) -- (-0.1, 1.25) -- (-0.1, .5);
		\draw [rounded corners] (1.1, .5) -- (1.1, -.2) -- (-.3, -.2) -- (1.1, 1.25) -- (1.1, .5);
		\draw (1, 1) -- (1, 0);
		\end{tikzpicture}
		\caption{A hypergraph with vertices $0, 1, 2$ and $3$ with two edges $(0, 3)$ and $(1, 3)$ as well as two hyperedges $(0, 2, 3)$ and $(1, 2, 3)$.}
		\label{example_hypergraph}
	\end{subfigure} 
	~
	\begin{subfigure}{0.4\textwidth}
		\centering 
		\begin{tikzpicture}[scale = 1]
		\draw (0, 0) -- (3, 0);
		\draw (0, 1) -- (3, 1);
		\draw (0, 2) -- (3, 2);
		\draw (0, 3) -- (3, 3);
		\node at (-.4, 0) {$\ket{+}$};
		\node at (-.4, 1) {$\ket{+}$};
		\node at (-.4, 2) {$\ket{+}$};
		\node at (-.4, 3) {$\ket{+}$};
		\node at (4, 0) {$0$};
		\node at (4, 1) {$1$};
		\node at (4, 2) {$2$};
		\node at (4, 3) {$3$};
		\draw[fill] (1, 3) circle [radius = 2pt];
		\draw[fill] (1, 0) circle [radius = 2pt];
		\draw[fill] (1, 2) circle [radius = 2pt];
		\draw (1, 0) -- (1, 3);
		\draw[fill] (1.5, 2) circle [radius = 2pt];
		\draw[fill] (1.5, 1) circle [radius = 2pt];
		\draw[fill] (1.5, 3) circle [radius = 2pt];
		\draw (1.5, 1) -- (1.5, 3);
		\draw[fill] (2, 0) circle [radius = 2pt];
		\draw[fill] (2, 3) circle [radius = 2pt];
		\draw (2, 0) -- (2,3);
		\draw (2.5, 1) -- (2.5, 2);
		\draw[fill] (2.5, 1) circle [radius = 2pt];
		\draw[fill] (2.5, 2) circle [radius = 2pt];
		\end{tikzpicture}
		\caption{Quantum circuit for generating the hypergraph state corresponding to the hypergraph depicted in figure \ref{example_hypergraph}.}
		\label{example_circuit}
	\end{subfigure}
	\caption{A hypergraph and its corresponding quantum circuit}
\end{figure}
The hypergraph states are a generalization of graph states or cluster states \cite{nielsen2006cluster}. If a hypergraph has $n$ vertices then the corresponding 
hypergraph state is an $n$-qubit state \cite{dutta2019permutation} belonging to $\mathcal{H}_2^{\otimes n}$. To construct a hypergraph state we first assign a qubit $\ket{+} = \frac{\ket{0} + \ket{1}}{\sqrt{2}}$, corresponding to every vertex. Also, for every hyperedge $\{u_1, u_2, \dots u_k\}$ we apply a $k$-qubit controlled $Z$ gate on the qubits corresponding to the vertices $v_1, v_2, \dots v_k$. These states can be expressed as 
\begin{equation}
\ket{G} = \frac{1}{\sqrt{2^n}} \sum_{i = 0}^{2^n - 1} (-1)^{f(\bin(i))}\ket{\bin(i)},
\end{equation} 
where $f: \{0, 1\}^n \rightarrow \{0, 1\}$ is a Boolean function with $n$ variables acting on the $n$-bit binary representation $\bin(i)$ of $i$. Clearly, the size of the state vector is $2^n$, where $n$ is the number of vertices in the hypergraph. For simplicity, we denote $2^n = N$, from now on.

The hypergraph in figure \ref{example_hypergraph} has four vertices. To every vertex, we assign a $\ket{+}$ state. Then we apply different controlled $Z$ operations on the qubits. For example, there is a  hyperedge $(0, 2, 3)$. Hence, we apply a $3$-qubit controlled $Z$ gate on $0$-th, $2$-nd and $3$-rd qubits. All the controlled $Z$ operations are depicted in figure \ref{example_circuit}, as a quantum circuit. The corresponding hypergraph state is a four-qubit state,
\begin{equation}\label{example_multiqubit_hyperrgaph_state}
\begin{split}
\ket{G} = \frac{1}{4} & [\ket{0000} + \ket{0001} + \ket{0010} + \ket{0011} + \ket{0100} + \ket{0101} - \ket{0110} + \ket{0111} \\
& + \ket{1000} - \ket{1001} + \ket{1010} + \ket{1011} + \ket{1100} - \ket{1101} - \ket{1110} + \ket{1111}],
\end{split}
\end{equation} 
where $\ket{0}$ and $\ket{1}$ represent the qubits. 

Note that, the set of vectors $\{\ket{\bin(i)}: i = 0, 1, 2, \dots (N - 1)\}$ forms a basis of  $\mathbb{C}^N$. We assume that the space is spanned by the set of $N$ dimensional vectors $\ket{i}$ for $i = 0, 1, 2, \dots (N - 1)$. Numerically, $\ket{i}$ is equivalent to $\ket{\bin(i)}$. To make our notations simplified we write $f(\bin(i)) = g(i)$ where $g : \{0, 1, 2, \dots (N - 1)\} \rightarrow \{0, 1\}$, which reflects the combinatorial structure of $G$. Therefore, corresponding to a hypergraph with $n$ vertices there is a qudit state of dimension $N$ in $\mathcal{H}_{N}$, as in equation (\ref{qudit_hypergraph_state}).


\section*{Acknowledgments}
SB acknowledges support from the Interdisciplinary Cyber-Physical Systems (ICPS) program of the Department of Science and Technology (DST), India through Grant No.: DST/ICPS/QuEST/Theme-1/2019/6. SB also acknowledges support from the Interdisciplinary Research Platform - Quantum Information and Computation (IDRP-QIC) at IIT Jodhpur.


\end{document}